\documentclass[aps,preprint]{revtex4}%

\pdfoutput=1

\usepackage{amsfonts}
\usepackage{amsmath}
\usepackage{amssymb}
\usepackage{graphicx}%

\begin{document}
\title{The solitons redistribution in Bose-Einstein condensate in quasiperiodic
optical lattice}
\author{G.N. Burlak$^{1}$, A.B. Klimov$^{2}$}
\affiliation{$^{1}$Center for Research on Engineering and Applied Sciences, Autonomous
State University of Morelos, Cuernavaca, Mor. 62210, Mexico. gburlak@uaem.mx,
$^{2}$Departamento de Fisica, Universidad de Guadalajara, Revolucion 1500,
Guadalajara, Jalisco, 44420, Mexico. klimov@cencar.udg.mx}
\begin{abstract}
We numerically study the dynamical excitations in Bose-Einstein condensate
(BEC) placed in periodic and quasi-periodic 2D optical lattice (OL). In case
of the repulsive mean-field interaction the BEC quantum tunnelling leads to a
progressive soliton's splitting and generating of secondary solitons, which
migrate to closest trapping potential minima. A nontrivial soliton dynamics
appears when a series of $\pi$-pulses (phase kicks) are applied to the optical
lattice. Such sudden perturbation produces a dynamic redistribution of the
secondary solitons, leading to a formation of an artificial solitonic
superlattice. Different geometries of OL are analyzed.

\end{abstract}
\maketitle

\section{Introduction}

Different types of solitons in Bose-Einstein condensates (BECs) have been
intensively studied both theoretically and experimentally
\cite{Alexander:2004a}-\cite{Zwierlein:2003a}. Recently, solitons supported by
weak attractive interactions between atoms were created in the condensate of
$^{7}Li$ trapped in strongly elongated traps\cite{Strecker:2002a},
\cite{Khaykovich:2002a}, \cite{Weber:2003a}. Ref. \cite{Kasprzak:2006a}
detailes a set of experiments showing evidence of the existence of BEC of
polaritons. Above a critical density, the authors observed massive occupation
of the ground state developing from a polariton gas at thermal equilibrium.

Also, a possibility to design $1D$ and $2D$ quantum systems by "freezing out"
one or two degrees of freedom by adding a $1D$ or $2D$ optical lattice to the
magnetic trap has been demonstrated experimentally \cite{Burger:2002a},
\cite{SabineStock:2005a} . However the soliton's dynamics in two dimensions is
much more involved and cannot be reduced to the one-dimensional case.

It is well known that solitonic solutions appearing in integrable models, such
as, for instance, $1D$ spatial solitons in the cubic nonlinear Schr\"{o}dinger
equation (NSE)\cite{Zakharov:1971a}, reveal particle-like behavior. In
particular, they remain unperturbed after collisions, completely preserving
their identities and form. However, there are no loosely bound stable solitons
in $2D$ (see e.g.\cite{Landau:1984a}).

The properties of solitons are well studied for a long optical lattice case
(gap solitons). In the repulsive condensate gap solitons (GSs) can be created
as a result of the interplay of the self-defocusing nonlinearity and periodic
potential induced by an optical lattice (OL) - an interference pattern created
by counterpropagating laser beams illuminating the condensate,
\cite{Abdullaev:2001a}-\cite{Gomez-Gardenes:2006a}.

The dynamics of BEC in $2D$ optical lattice are currently a subject of
intensive theoretical and experimental studies\cite{Baizakov:2004a}%
-\cite{Rychtarik:2004a}. An interesting direction consists in using trapping
quasiperiodic potentials for generation of several nonlinear excitations. The
authors \cite{HidetsuguSakaguchi:2006a} have constructed some families of
solitons for one- and two-dimensional Gross-Pitaevskii equations with a
repulsive nonlinearity and a potential of quasicrystallic type (in the $2D$
case, the potential corresponds to a fivefold optical lattice). Some stable
$2D$ and $3D$ solitons in the self-attractive Gross-Pitaevskii equation were
found in \cite{Baizakov:2004a} by using the variational approximation and
direct simulations in real and imaginary time. Also, investigation of
stability of localized vortices in an effectively $2D$ (\textquotedblleft
pancake-shaped\textquotedblright) trapped BEC with negative scattering length
was performed in \cite{DumitruMihalache:2006a}.

Usually, in experiments on solitons generation in BEC it is possible to obtain
a unique soliton which is stabilized in gap zones of a periodic OL. However
from the point of view of quantum-information technologies the possibility of
generation of nonlinear excitations with complex structure, similarly to
multisoliton state is more important.\textbf{ }Several nonlinear regimes of
interaction of such excitations have been studied in the literature.
Nevertheless, the problem of experimental generation of multisolitonic states
in BEC is still far from being completely solved.

In this paper we explore an intermediate case when the size of BEC is about
ten (or less) spatial periods of the optical lattice. In this situation the
approximation of a perfect potential periodicity, $V(\mathbf{x+L}%
)=V(\mathbf{x})$ (necessary for the soliton stabilization) strictly speaking,
is not valid, and a more realistic model should be constructed. In such
configurations the dynamics of a single soliton are defined by the structure
of the trapping potential in the vicinity of the soliton, rather by the far OL
order. One can say that a soliton is subjected to a local spatially
non-uniform nonlinear force field, which is defined by OL configuration, and
by the excitation's amplitude. Such a field drastically changes the soliton
dynamics: the soliton can tunnel into the closest minima (in the case of
repulsive interaction), or turn to a self-compression dynamics (in the case of
attractive interaction).

We focus on the repulsive interaction case. Then, if the initial soliton
position does not coincide with the minima of OL, the shape of such excitation
will be split into fragments tending to flow into the closest minima of
trapping potential, and the soliton dynamics essentially depends on a local
spatial structure of OL.

This observation suggests the exploration the soliton evolution in BEC when
the spatial OL configuration is rapidly changed (which can be easily achieved
experimentally). We have found that remarkable dynamics occur in situation
when a soliton, steady in an initial OL configuration, becomes nonequilibrium
in a new OL. This establishes the possibility to drastically increase the
speed of forming the secondary solitons with such OL switching. Below we
demonstrate that as a result of such a dynamics an artificial solitonic
superlattice in BEC is formed.

In this paper we analyze dynamics of a soliton placed in periodic or\textbf{
}quasi-periodic optical lattice\textbf{s} in $2D$ Bose-Einstein condensate
(BEC) in the intermediate case (when the spatial scale of the condensate area
is about $10$ or less OL wavelengths), when the OL is suddenly switched after
applying periodical $\pi$ phase shifts (phase kick). It will be shown that if
the period of switching is sufficiently large (with respect to the tunneling
time) secondary solitons are formed in OL minima. At later (successive) $\pi$
-switching such secondary solitons pass through series of metastable states
and migrate from the center to the periphery. In a suitable moment this
process can be interrupted, which allows a free interaction of generated
nonlinear excitations in agreement with the fundamental $2D$ nonlinear
Schr\"{o}dinger equation.

The paper is organized as follows. In Section II we discuss basic equations
for BEC atoms placed into OL. In Section III we present numerical studies of
the soliton dynamics in the repulsive mean-field interaction for $2D$ case
with periodical $\pi$-phase OL shifts. Various geometries of OL, including
periodic and quasiperiodic, are analyzed. We also discuss some peculiarities
of the evolution in the attractive interaction case. In the last Section we
summarize our results.

\section{Basic equations}

In order to trap a Bose-Einstein condensate in a \textbf{(}quasi\textbf{)}%
-periodic potential, it is sufficient to exploit the interference pattern
created by two or more overlapping laser beams and the light force exerted on
the condensate atoms. We mainly focus on the physical situation when the
number of atoms is sufficiently large, so that atomic number fluctuations are
negligible and the mean-field approximation can be applied. In this approach
an anisotropic BEC cloud, loaded into a two-dimensional optical lattice
potential $V(\mathbf{r})$, is described by the Gross-Pitaevskii (GP) equation
($T=0$)%

\begin{equation}
i\dfrac{\partial\Phi}{\partial\tau}=[-\triangledown_{\perp}^{2}+V(\mathbf{r}%
)\mathbf{+}G\left\vert \Phi\right\vert ^{2}]\Phi\text{,} \label{GPE}%
\end{equation}
where $\mathbf{r}=\{x,y\}$, $\mathbf{\nabla}_{\bot}=\{\partial_{x}%
,\partial_{y}\}$, $\Phi$ is the condensate wave function, $G=+1$, $-1$ for
repulsive and attractive interaction correspondingly. This equation is
obtained by assuming a tight confinement in the direction perpendicular to the
lattice ({"}pancake{"} trapping geometry) and the standard dimensionality
reduction procedure (see e.g.\cite{FrancoDalfovo:1999a},
\cite{OliverMorsch:2006a}). According to Eq.(\ref{GPE}), a particle with the
(condensate) wave function $\Phi(\mathbf{r},t)$ evolves in the external
potential $V(\mathbf{r})$ induced by OL plus the mean-field potential created
by the remaining particles.

It is made dimensionless by using the characteristic length $h_{0}%
=p/\sqrt{8\pi a_{S}}$, energy $E_{0}=g/p^{2}=\hbar/t_{0}$, and time
$t_{0}=\hbar p^{2}/g=mp^{2}/4\pi\hbar a_{S}$, here and below $m$ and $N_{0}$
are the mass and number of the trapped atoms respectively, $g=4\pi\hbar
^{2}a_{s}/m$ is the interaction strength, and $a_{s}$ is the $s$ -wave
scattering length. The wave function is made dimensionless as $\Phi\rightarrow
p\Phi$, where the factor $p$ is specified by the normalization conditions
$N_{0}=\int\left\vert p\Phi\right\vert ^{2}dV_{0}$ (i.e., $n_{0}%
(\mathbf{r,\tau})=\left\vert p\Phi(\mathbf{r,\tau})\right\vert ^{2}$), where
$V_{0}$ is the volume of the condensate.

\begin{figure}
[ptb]
\begin{center}
\includegraphics[
natheight=4.375100in,
natwidth=5.833200in,
height=4.4261in,
width=5.892in
]%
{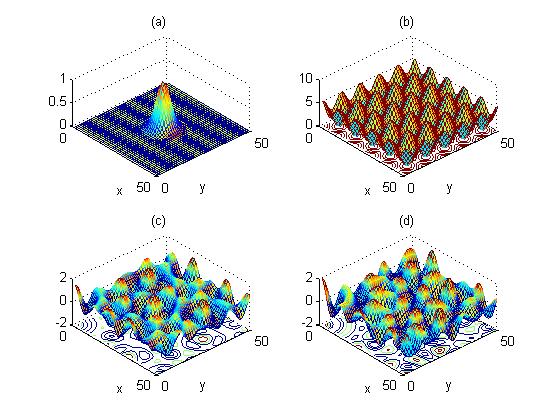}%
\caption{(a) Init Gaussian package. (b),(c),(d) two-dimensional distribution
of the trapping potential $V(x,y,0)$ induced by OL, Eq.(\ref{OptLatticeV}) for
cases (b) $N=4$; (c) $N=5$; (d) $N=7$.}%
\label{Pic_lattinit200x200(n457)}%
\end{center}
\end{figure}

We consider a quasiperiodic \textit{time-dependent} trapping potential $V$ of
the following form,%

\begin{equation}
V=V(x,y,\tau)=V_{c}+\varepsilon\sum\limits_{n=1}^{N}\cos(\mathbf{k}%
^{(n)}\mathbf{r+\pi\theta(\tau))=}V_{c}+\varepsilon(1-2\mathbf{\theta(\tau
)})\sum\limits_{n=1}^{N}\cos(\mathbf{k}^{(n)}\mathbf{r}\mathbf{)}%
\text{,}\label{OptLatticeV}%
\end{equation}
where $\mathbf{k}^{(n)}=\{\cos(2\pi(n-1)/N,\sin(2\pi(n-1)/N\}$, $V_{c}=const$,
and the time-dependence appears through the pulse function $\mathbf{\theta
(\tau)=1}$ for $(2n+1)T_{t}<\tau$ $\leq(2n+2)T_{t}$ and $\mathbf{\theta
(\tau)=0}$ otherwise, $T_{t}$ \ is a time period, $n=0,1,2...$. For stationary
OL ($T_{t}\rightarrow\infty)$ and $N=5$ such a potential case was studied in
\cite{HidetsuguSakaguchi:2006a}. Some typical time independent potentials,
$V(x,y,0)$, are shown in Figs.\ref{Pic_lattinit200x200(n457)} (b),(c),(d) for
the $N=4,5,7$ cases correspondingly. Such quasiperiodic optical lattices can
be created in the $2D$ case, as a combination of $N=4$, $N=5$ (Penrose tiling
- a pattern of tiles, which completely cover an infinite plane in an aperiodic
manner) or $N\geq7$ quasi-$1D$ sublattices with wave vectors $\mathbf{k}%
^{(n)}$. The band-gap spectrum of $2D$ photonic crystals of the PT type has
been studied in \cite{Kaliteevski:2000a}, \cite{Kaliteevski:2004a},
\cite{MasashiHase:2002a}\textbf{. }Recently the interesting properties of such
lattices (e.g., Penrose lattices), have been discovered for quasiperiodic
pinning arrays\cite{Misko:2006a}).%

From now on we will concentrate on a repulsive interaction case, $G=1$. First
of all, we \ note that the generalized momentum density is not conserved
because the trapping potential $V(x,y)$ induces a spatial inhomogeneity in a
system \cite{Goldstein:2002a}. Let us estimate the force acting on the
condensate particles. It is well known that the GP equation (\ref{GPE}) can be
generated from the Lagrangian density%

\begin{equation}
L=\dfrac{i}{2}[\Phi^{\ast}\Phi_{\tau}-\Phi\Phi_{\tau}^{\ast}]-\left\vert
\triangledown_{\perp}\Phi\right\vert ^{2}-V(\mathbf{r})\left\vert
\Phi\right\vert ^{2}-\dfrac{G}{2}\left\vert \Phi\right\vert ^{4}\text{,}
\label{Lagr}%
\end{equation}
and the corresponding full condensate energy has the form%

\begin{equation}
E=\int d\mathbf{r}[\left\vert \mathbf{\triangledown}_{\perp}\Phi\right\vert
^{2}+V(r)\left\vert \Phi\right\vert ^{2}+\dfrac{G}{2}\left\vert \Phi
\right\vert ^{4}]\mathbf{=}const\text{.} \label{EnergyGP}%
\end{equation}
Observe that the constant part of the trapping potential, $V_{c}$, can be
removed from the Lagrangian (\ref{Lagr}) by renormalizing the wave function as follows,%

\begin{equation}
\Phi\rightarrow\Phi\exp(-iV_{c}\tau)\text{.} \label{new Phi}%
\end{equation}
Thus, without loss of generality we suppose that the potential $V(x,y)$ in
(\ref{GPE}) and (\ref{Lagr}) can be renormalized such that $V_{min}(x,y)=0$
and $V(x,y)\geq0$. Note that in the case of such renormalization the energy
(\ref{EnergyGP}) is shifted by $V_{c}N_{0}$, which allows us to associate
$V_{c}$ with the chemical potential $\mu=dE/dN_{0}$.

The Lagrangian (\ref{Lagr}) generates the following force field acting on the
soliton,\hspace{5mm}%

\begin{equation}
F(x,y)=\dfrac{\delta L}{\delta\Phi^{\ast}}=\dfrac{\partial L}{\partial
\Phi^{\ast}}-\triangledown\dfrac{\partial L}{\partial\triangledown\Phi^{\ast}%
}=F_{V}+F_{S}+F_{N}\text{,} \label{SolForce}%
\end{equation}
where%

\begin{equation}
F_{V}=-V(x,y,\tau)\Phi\text{, }F_{N}=-G\left\vert \Phi\right\vert ^{2}%
\Phi\text{, }F_{S}=\triangledown_{\perp}^{2}\Phi\text{.} \label{SolForceAll}%
\end{equation}
One can observe that $F_{V}=0$ in the vicinity of $V_{min}=0$ and therefore in
the potential minima the soliton evolution is governed by the standard
nonlinear Schr\"{o}dinger model dynamics, i.e. only subjected to the forces
$F_{S}$ and $F_{N}$. This, in particular, leads to a very low tunneling rate
even for an intermediate depth of OL. Thus, as we have numerically
corroborated, a sufficiently narrow soliton placed in one of the potential
minima is practically a steady-state.

It is clear that a deviation from $V_{min}=0$ generates a nonzero $F_{V},$
which essentially modifies the soliton behavior. The effect of $F_{V}$ becomes
predominant in the vicinity of the maximum of the potential. Thus, after
applying a phase kick (when the minima of the potential are suddenly converted
into maxima), the initial "steady-state" soliton starts to rapidly evolve in
the direction of the closest minima of the new potential. Besides, in the
vicinity of the soliton maximum (where $\Phi$ is real and positive), the
forces $F_{N}$ and $F_{S}$ are all negative. These forces stretch and deform
the initial packet leading to the soliton splitting. It is worth noting that
due to the energy conservation (\ref{EnergyGP}), the decay of some part of the
BEC soliton amplitude $\left\vert \Phi\right\vert $\ must be compensated by
sharp gradients of some other parts.

\section{Numerical results}

\subsection{Repulsive interaction}

\begin{figure}
[ptb]
\begin{center}
\includegraphics[
natheight=4.375100in,
natwidth=5.833200in,
height=4.4261in,
width=5.892in
]%
{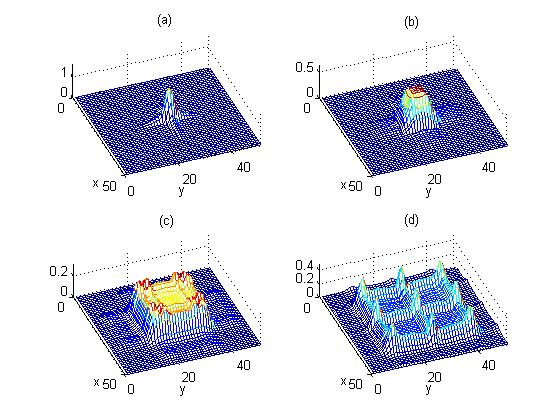}%
\caption{Evolution of BEC soliton for period $\pi$-phase shift $T_{t}=50$ and
different times: (a) $\tau=40$; (b) $\tau=80$; (c) $\tau=120$; (d) $\tau=160$.
See details in text.}%
\label{Pic_200x200g1}%
\end{center}
\end{figure}

\begin{figure}
[ptb]
\begin{center}
\includegraphics[
natheight=4.375100in,
natwidth=5.833200in,
height=4.4261in,
width=5.892in
]%
{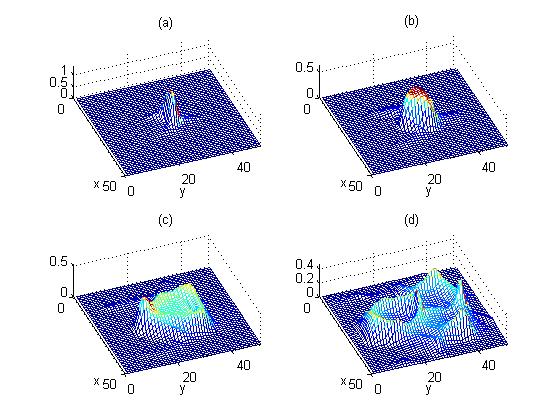}%
\caption{The same as in Fig.\ref{Pic_200x200g1}, except $N=5$ in trapping
potential $V(x,y,\tau)$ Eq.(\ref{OptLatticeV}). }%
\label{Pic_200x200n5g1}%
\end{center}
\end{figure}

\begin{figure}
[ptb]
\begin{center}
\includegraphics[
natheight=4.375100in,
natwidth=5.833200in,
height=4.4261in,
width=5.892in
]%
{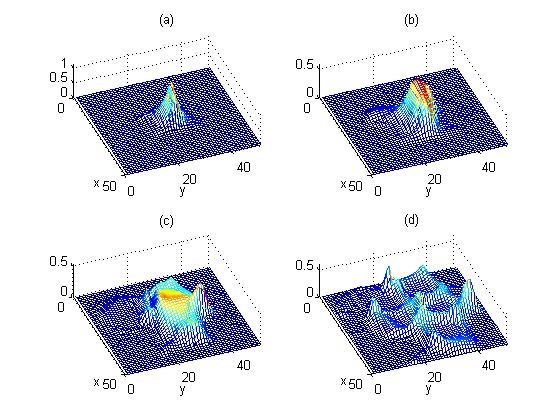}%
\caption{The same as in Fig.\ref{Pic_200x200g1}, except $N=7$ in trapping
potential $V(x,y,\tau)$ Eq.(\ref{OptLatticeV}). }%
\label{Pic_200x200n7g1}%
\end{center}
\end{figure}

\begin{figure}
[ptb]
\begin{center}
\includegraphics[
natheight=4.375100in,
natwidth=5.833200in,
height=4.4261in,
width=5.892in
]%
{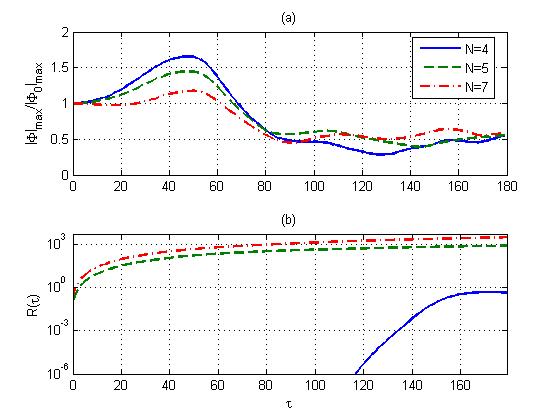}%
\caption{Time dynamics in case of repulsive interaction ($G=1$). (a)
$\left\vert \Phi_{\max}(\tau)\right\vert /\left\vert \Phi_{\max}(0)\right\vert
$, and (b) mass center $R(\tau)$, see (\ref{dipoleMom}). Solid line, dash line
and dash-dots line correspond to $N=4$, $N=5$ and $N=7$ cases accordingly for
trapping potential $V(x,y,\tau)$ in Eq.(\ref{OptLatticeV}). Here $\tau\equiv
n$ (n is the time step index).}%
\label{Pic_maxu200x200g1}%
\end{center}
\end{figure}

Previous arguments have only heuristic meaning and do not provide a
quantitative description of the soliton dynamics in the $2D$ case, especially
for time-dependent OL potential. The deeper insight into the BEC dynamics in a
general OL depth case can be reached only by numerical methods. We have
investigated two-dimensional dynamics for various initial conditions and
different parameters of the trapping potential. The results are shown in
Figs.\ref{Pic_200x200g1}-\ref{Pic_200x200g_1}. For our numerical experiments
we used the combination of known splitting methods\cite{Press:2002a}
generalized to the nonlinear case. We have numerically solved Eq.(\ref{GPE})
in domain $200\times200$ with zero boundary conditions for times $\tau<200$.
Greater temporal intervals normally require a spatial grid with greater size.
Our calculations were done in the complex $\Phi$\ plane to observe the
evolution of both amplitude and phase of the soliton. Both the norm of the
wave function and energy (\ref{EnergyGP}) were conserved with good accuracy
(less then $1\%$). We have studied the evolution of the initial Gaussian
packet, Fig.\ref{Pic_lattinit200x200(n457)}(a),%

\begin{equation}
\Phi(x,y,0)=\Phi_{0}\exp[-q_{x}\left(  x-x_{0}\right)  ^{2}-q_{y}\left(
y-y_{0}\right)  ^{2}) \label{Init state}%
\end{equation}
for different factors $q_{x,y}$, and the amplitude $\Phi_{0}$, which is a
function of the total number of trapped atoms $N_{0}$. We have numerically
analyzed the packet (\ref{Init state}) evolution initially centered at
different points of the plane $(x_{0}$,$y_{0})$, corresponding either to
minima ($V_{min}$) and maxima ($V_{max}$) of the potential.

Let us start with a periodic trapping potential Eq.(\ref{OptLatticeV}),
corresponding to $N=4$ (see Fig.\ref{Pic_lattinit200x200(n457)} (b)) and for
$G=1$. We have used the parameters: $q_{x}=q_{y}=$\ $0.004$, $\varepsilon
=0.7$, $\Phi_{0}=1$, and the initial packet was placed at the central minimum,
Fig.\ref{Pic_lattinit200x200(n457)}(a). In the case of a stationary potential
the initial packet evolves in a well known way by splitting into fragments and
slowly tunneling into the closest minima of the optical lattice. Nontrivial
dynamics arise after sudden (short with respect to the scale $t_{0}$) shift by
$\pi$ of the optical lattice phase, when OL minima are switched with their
maxima. Being placed in these new maxima the soliton flow out into new closest
minima and after a certain transition time the secondary solitons are
generated in those minima. In our simulations the period of the phase shift,
$T_{t}=50$, is chosen larger than the above mentioned transition time
(proportional to the characteristic time of quantum tunneling), so that the
details of the initial state are washed out and the secondary "solitons"
acquire well pronounced shape.

These dynamics are shown in Fig.\ref{Pic_200x200g1}. In
Fig.\ref{Pic_200x200g1} (a) the soliton state at time $\tau=40$ before the
first phase switching is shown. One can observe the beginning of slow
condensate tunneling into closest potential minima. In Fig.\ref{Pic_200x200g1}
(b) the condensate state at $\tau=80$ (after first phase switching) is
presented. One can appreciate an essential reduction and widening of pulse
amplitude. We observe that such dynamics become significantly faster with
respect to the stationary case\textbf{.} At $\tau=120$ (after second phase
switching), Fig.\ref{Pic_200x200g1} (c), the state acquires a more complex
shape with quite pronounced secondary solitonic peaks at the corners. The
final state at $\tau=160$ (after third phase switching) is shown in
Fig.\ref{Pic_200x200g1} (d). We observe already a well formed artificial
solitonic superlattice with the peaks located in the trapping potential
minima. It is clear that the details of this new soliton structure depend
mainly on the local vicinity of OL.%

\ 

Due to a recent interest to quasiperiodic structures, we also analyze the BEC
evolution\ in an optical structure with a lower symmetry trapping potential.
Such dynamic is shown in Fig.\ref{Pic_200x200n5g1} for potential
(\ref{OptLatticeV}) with $N=5$ and in Fig.\ref{Pic_200x200n7g1} for potential
(\ref{OptLatticeV}) with $N=7$ (see Fig.\ref{Pic_lattinit200x200(n457)} (c)
and (d) accordingly). The case $N=5$ is essentially a two-dimensional
Penrose-tiled lattice, where the tiles are two kinds of rhombus: a thin tile
(with vertex angles of $36%
{{}^\circ}%
$ and $144%
{{}^\circ}%
$) and a fat tile ($72%
{{}^\circ}%
$ and $108%
{{}^\circ}%
$)\cite{Penrose:1979a}. We observe from Fig.\ref{Pic_200x200n5g1} and
Fig.\ref{Pic_200x200n7g1} that for quasiperiodic potentials the system evolves
in a less symmetric way. The condensate is distributed on closed smooth cells
without strongly pronounced solitonic peaks. Such cells appear more smooth in
the $5$-fold case, see Fig.\ref{Pic_200x200n5g1}. This effect is similar to
$2D$ Penrose lattices in superconductors, where the pinning of vortices is
related to matching conditions between the vortex lattice and the
quasiperiodic lattice of pinning centers\cite{Misko:2006a}. We notice again
that the final structure of a condensate still has the form of localized
states distributed on the minima of the potential.%

\bigskip%

It is representative to illustrate the state evolution as the time behavior of
the amplitude $\left\vert \Phi\right\vert $ of the maximal peak, as shown in
Fig.\ref{Pic_maxu200x200g1}(a). It can be observed that the amplitude after
some period of growing (which is accompanied by the stretching of the initial
distribution due to non-linear effects) starts to diminish and reaches its
minimum (quasistationary value) at $\tau=130$, although the splitting is still continuing.%

To evaluate quantitatively the behavior of BEC shape we have studied the
behavior of the soliton "center of mass" $\mathbf{R}(\tau)$ defined as
\begin{equation}
\mathbf{R}(\tau)=\int(\mathbf{r}-\mathbf{r}_{0})\left\vert \Phi(\mathbf{r}%
,\tau)\right\vert ^{2}d\mathbf{r}\text{,} \label{dipoleMom}%
\end{equation}
where $\mathbf{r}_{0}$\ indicates the initial soliton position. The dynamics
of $R(\tau)=\left\vert \mathbf{R}(\tau)\right\vert $\ are shown in
Fig.\ref{Pic_maxu200x200g1}(b). One can observe that $R(\tau)$ grows in a very
different way for periodic and quasiperiodic potentials,\textbf{ }which is a
reflection of the different local structure of OL. On the other hand, while
the minima are progressively filling up with the condensate the quantity
$R(\tau)$ tends to saturation in both cases.

\subsection{Attractive interaction.}

Although the main goal of this paper is studying the repulsive interaction,
for completeness, we also discuss the dynamics of the initial Gaussian pulse
(\ref{Init state}) in the attractive interaction case ($G=-1$). Let us recall
that for the $2D$ nonlinear Schr\"{o}dinger model (without potential) the
dynamics of the average radius of an excitation are (see e.g. Chap. VIII in
Ref.\cite{Landau:1984a}) $\left\langle R\left(  \tau\right)  \right\rangle
^{2}=\int\rho^{2}\left\vert \Phi\right\vert ^{2}d^{2}r\simeq(E/N_{0})\tau
^{2}+R_{0}^{2}$ (see also\cite{CatherineSulem:1999a} and references therein).
So, if the energy (\ref{EnergyGP}) is negative, $E<0$, then such an excitation
undergoes a self-compressing. For the initial Gaussian (\ref{Init state}) this
yields the inequality $\Phi_{0}^{2}>q_{\mathbf{x,y}}$, which is satisfied in
our case. However a similar analytical condition for $2D$ BEC with OL is not
established yet in the literature. Therefore we have studied the soliton
evolution for the 2D attractive interaction numerically for the initial state
Fig.\ref{Pic_lattinit200x200(n457)} (a) and different OL geometries
(Fig.\ref{Pic_lattinit200x200(n457)}(b,c,d)), similar to the repulsive case.

\begin{figure}
[ptb]
\begin{center}
\includegraphics[
natheight=4.375100in,
natwidth=5.833200in,
height=4.4261in,
width=5.892in
]%
{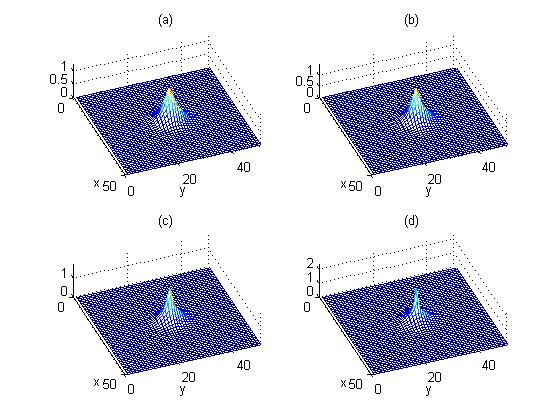}%
\caption{Self-compressing of initial Gaussian package at attractive
interaction $G=-1$ and the period of $\pi$-phase shift $T_{t}=8$. (a)
$\tau=15$; (b) $20$; (c) $25$; (d) $30.$}%
\label{Pic_200x200g_1}%
\end{center}
\end{figure}

As it can be appreciated from Fig.\ref{Pic_200x200g_1} the soliton dynamics
completely change in the attractive case \ $G=-1$. It is well known that for
the $2D$ attractive interaction the collapse dynamics takes place and there
are no loosely bound stable states. We have found that for our initial state
the collapse dynamics take place and there are no loosely bound stable states
in two dimensions. Nevertheless, it turns out that such instabilities can be
exploited in order to prepare BEC solitons\cite{OliverMorsch:2006a}. We have
investigated such a dynamics for various geometries of the trapping potential.
In contrast to the repulsive interaction case we found that the
self-compressing behavior is practically insensible to changing OL symmetry.
We observe from Fig.\ref{Pic_200x200g_1} that just a single pulse is generated
from the initial packet, which evolves without any splitting or visible
deformation. Such self-compression leads to a singularity and therefore, the
local structure of OL in the vicinity of the soliton does not notably affect
its evolution. Obviously, the tunneling processes quickly become practically
negligible in this case. We have also found that small shifts of the soliton
center from the minimum of the potential lead to the same dynamics, and in
particular, do not change the single-peak structure of the final state.
Generating of two and more peaks as well as stabilization of self-compressing
can be reached via a specific initial state and/or OL geometry. However in
general this question remains open and will be studied elsewhere.

\bigskip%

\section{Conclusion}

We have studied the dynamics of soliton in Bose-Einstein condensate (BEC)
placed in periodic or quasi-periodic $2D$ optical lattices (OL). In the case
of the repulsive mean-field interaction the BEC tunneling leads to progressive
deformation and splitting of solitons. It results in the generation of
secondary solitons migrating to closest trapping potential minima. We found
that nontrivial dynamics arise when a series of $\pi$-phase shifts (phase
kicks) are applied to the OL, so that the minima and maxima of OL are
periodically exchanged. This results in a progressive migration of the
solitons from the initial position to the periphery with the forming of the
artificial solitonic superlattice. We have found that the geometry of forming
solitonic superlattice is rather sensitive to the geometry of the trapping potential.

Such an effect suggests an effective and simple method for splitting of the
initial BEC soliton and generation of multisolitonic nonlinear excitations in
the experiment, and thus\textbf{, }can be, in principle, applied in quantum
information technology, since it allows a creation and manipulation of
solitons as bits (qubits) in BEC quantum structures.

\section{Acknowledgements}

This work of G.B is partially supported by CONACyT grant 47220. The work of
A.K. is partially supported by CONACyT grant 45704.

\end{document}